# Hybrid Quantum-Classical Simulations of Graphene Analogues: Adsorption Energetics Beyond DFT

Archith Rayabharam[1], N. R. Aluru[1,*]

[1] Walker Department of Mechanical Engineering, Oden Institute for Computational Engineering and Sciences, The University of Texas at Austin, Austin, Texas, USA

*Corresponding author, email: aluru@utexas.edu

Abstract

**Understanding strongly correlated systems is essential for advancing quantum chemistry and materials science, yet conventional methods like Density Functional Theory (DFT) often fail to capture their complex electronic behavior. To address these limitations, we develop a hybrid quantum-classical framework that integrates Multiconfigurational Self-Consistent Field (MCSCF) with the Variational Quantum Eigensolver (VQE). Our initial benchmarks on water dissociation enabled the systematic optimization of key computational parameters, including ansatz selection, active space construction, and error mitigation. Building on this, we extend our approach to investigate the interactions between graphene analogues and water, demonstrating that our framework produces binding energies consistent with high-accuracy quantum methods. Furthermore, we apply this methodology to predict the binding energies of transition metals (Fe, Co, Ni) on both pristine and defective graphene analogues, revealing strong charge transfer effects and pronounced multireference character—phenomena often misrepresented by standard DFT. In contrast to many existing quantum algorithms that are constrained to small molecular systems, our framework achieves chemically accurate predictions for larger, strongly correlated systems such as metal–graphene complexes. This advancement highlights the capacity of hybrid quantum-classical approaches to address complex electronic interactions and demonstrates a practical route toward realizing quantum advantage for real-world materials applications in the Noisy Intermediate-Scale Quantum (NISQ) era.**

Introduction

Quantum simulations provide a critical pathway for understanding the behavior of complex molecular and material systems by leveraging the full formalism of quantum mechanics. While DFT[1] remains a cornerstone for electronic structure calculations, it often fails to capture strong electron correlations, multireference effects, and long-range dispersion interactions—features essential to accurately modeling transition metal complexes, 2D materials, and catalytically active sites. Post-Hartree-Fock methods such as Configuration Interaction (CI)[2,3] and Coupled Cluster (CC)[4] offer improved accuracy but are prohibitively expensive for systems beyond a modest number of electrons or orbitals.

Quantum computing offers a fundamentally new paradigm to overcome these challenges. By encoding quantum states directly onto qubits, quantum algorithms promise polynomial or even exponential advantages in simulating many-body electronic systems. Among the most promising methods for near-term devices is VQE[5], a hybrid quantum-classical algorithm designed for NISQ hardware. VQE and related methods have been successfully implemented for small molecular systems using superconducting qubits[6, 7] and trapped-ion platforms, and are increasingly supported by theoretical advancements in qubit encoding, error mitigation, and ansatz optimization[7-12]. Quantum computations have also been applied to bond dissociation studies[7-9, 11, 13-16], offering insights into reaction and material failure mechanisms and electronic structure behavior.

Strongly correlated materials (SCMs)—such as transition metal oxides, low-dimensional carbon systems, and molecular magnets—remain a major frontier in quantum chemistry and condensed matter physics[17-21]. Their electronic structure is dominated by multireference and entanglement effects that defy mean-field approximations. While quantum algorithms like Quantum Phase Estimation[22] offer asymptotically exact solutions, their depth and coherence-time requirements render them impractical on current quantum hardware[23, 24]. Hybrid quantum-classical methods[25], especially those that combine multiconfigurational classical frameworks with quantum solvers, offer a tractable alternative for studying these systems today.

In this work, we develop a hybrid quantum-classical framework that integrates MCSCF theory with VQE to accurately model strongly correlated systems beyond the capabilities of traditional methods. Unlike many existing quantum algorithms limited to minimal basis sets or small molecules, our approach delivers chemically accurate predictions for complex, extended systems such as transition metal–graphene analogues. We first validate the framework on water dissociation, optimizing key parameters including ansatz selection, active space configuration, and classical optimization strategies. We then apply it to compute the binding energies of water and transition metals (Fe, Co, Ni) on both pristine and defective graphene analogues. Our results capture strong charge-transfer effects and multi-reference electronic character that standard DFT fails to resolve. These systems exemplify strongly correlated behavior and represent scientifically and technologically important problems—such as catalysis, sensing, and materials design—that serve as rigorous stress tests for electronic structure methods. Our findings highlight the growing utility of quantum-classical workflows in addressing these challenges and underscore their relevance to real-world applications in the NISQ era.

**Fragmentation based approaches: MCSCF-VQE**

The limitations of current quantum hardware have sparked interest in divide-and-conquer strategies for practical quantum computations. Many real-world systems are too complex for direct solutions of the Schrödinger equation, necessitating approximate yet accurate quantum methods. By partitioning large problems into smaller, independently solvable subproblems, this approach reduces computational cost and suits the constraints of NISQ devices. Methods like Density Matrix Embedding Theory (DMET)[26] and Quantum Defect Embedding Theory[27] exemplify this strategy, enabling efficient and scalable simulations of quantum materials. However, none have tackled adsorbate interactions with graphene analogues or the multireference aspect of defect sites, which is the focus of this work. We use a fragmentation-based framework that combines VQE with post-Hartree-Fock methods (e.g., CC) to efficiently compute ground-state energies and potential energy surfaces in strongly correlated materials. This approach leverages quantum computing to bridge theoretical modeling and practical material design. The workflow consists of five steps (Figs. 1a, 1c):

i. System Definition and Active Space Selection
   ii. Construction of the Effective Hamiltonian
   iii. Qubit Mapping
   iv. Quantum State Preparation
   v. Optimization

**Step (i): System Definition and Active Space Selection.** The first step in our workflow is system definition and active space selection, involving the specification of atomic structure, basis set, spin, and charge. To efficiently capture electron correlation, the system is partitioned into an active space—treated with high-level methods such as CC—and an environment approximated using mean-field approaches like HF or DFT. Various active space selection techniques have been proposed for materials systems[27, 28], but in most cases, electrons and orbitals near the Fermi level play a crucial role in defect-related properties and are thus prioritized for inclusion in the active space. This strategic partitioning ensures computational efficiency while accurately capturing the essential physics of the system. The active space selection for the framework was validated with $H_2O$ dissociation, computing the potential energy surface by varying the length of one O–H bond while fixing the other.

**Step (ii): Constructing the Effective Hamiltonian.** The next step involves constructing the effective Hamiltonian of the chosen active space, which includes its most important electronic interactions. A second-quantized Hamiltonian is constructed based on the system parameters and is mathematically expressed as:

$$\widehat{H}_{elec} = \sum_{pq} h_{pq} \hat{a}_p^\dagger \hat{a}_q + \frac{1}{2} \sum_{pqrs} h_{pqrs} \hat{a}_p^\dagger \hat{a}_q^\dagger \hat{a}_r \hat{a}_s \tag{1}$$

where $\hat{a}_p^\dagger$ and $\hat{a}_p$ represent the creation and annihilation operators, respectively. The one-body integrals, which account for kinetic energy and nuclear attraction of electrons, are given by,

$$h_{pq} = \int \phi_p^*(r) \left( -\frac{1}{2} \nabla^2 - \sum_I \frac{Z_I}{R_I - r} \right) \phi_q(r) dr \tag{2}$$

where $\phi_p(r)$ represents the wavefunction of electron 'p', $Z_I$ is the atomic number of the nucleus 'I' and $R_I$ is the position of the nucleus 'I' while the two-body integrals, describing electron-electron interactions, are expressed as,

$$h_{pqrs} = \int \frac{\phi_p^*(r_1) \phi_q^*(r_2) \phi_r(r_2) \phi_s(r_1)}{|r_1 - r_2|} dr_1 dr_2 \tag{3}$$

These integrals capture the fundamental interactions governing the system's electronic structure, forming the basis for quantum simulations and electronic property calculations.

**Step (iii): Qubit Mapping.** To perform quantum computations, the fermionic Hamiltonian must be transformed into a qubit-compatible representation. This process, known as qubit mapping, encodes fermionic operators into spin operators using techniques such as the Jordan-Wigner transformation[29]. The Jordan-Wigner mapping ensures proper anti-commutation relations between fermionic states by representing creation and annihilation operators in terms of Pauli matrices:

$$\hat{a}_p^\dagger = \otimes_{p=1}^{q-1} Z_q \otimes (X_p - iY_p)$$
$$\hat{a}_p = \otimes_{p=1}^{q-1} Z_q \otimes (X_p + iY_p) \qquad (4)$$

where X, Y, Z are Pauli matrices[30, 31] and $\otimes$ represents the tensor product. The mapped qubit Hamiltonian can be expressed as follows,

$$\hat{H}_{qubit} = \hat{H} = \sum_i c_i \hat{P}_i \qquad (5)$$

where $\hat{P}_i$ are combinations of tensor products of Pauli matrices and $c_i$ are parameters dependent on $h_{pq}$ and $h_{pqrs}$

Once expressed in this form, the Hamiltonian can be implemented on quantum circuits, allowing for the evaluation of expectation values. Recent advancements in fermion-to-qubit mappings[32-40] have explored alternative transformations beyond Jordan-Wigner, optimizing qubit efficiency and reducing circuit complexity. These developments are crucial for improving the feasibility of simulating quantum systems on near-term quantum hardware.

**Step (iv): Quantum State Preparation.** In this step, a trial wavefunction (ansatz) is constructed by parameterizing the many-body quantum state through a quantum circuit. The ansatz is applied to an initial reference state $|\Psi_0\rangle$ using a unitary transformation $U_{\vec{\theta}} = U(\vec{\theta})$, where $\vec{\theta}$ represents a set of tunable parameters:

$$\Psi(\vec{\theta}) = \Psi_{\vec{\theta}} = U_{\vec{\theta}}|\Psi_0\rangle \qquad (6)$$

Once the quantum state is prepared, the system's energy can be estimated by executing the quantum circuit and measuring the expectation values. An illustration of such a circuit is shown in Fig. 1. (b).

**Step (v): Optimization.** A classical optimizer iteratively sums the measured expectation values and adjusts the parameters $\vec{\theta}$ to minimize the energy, ultimately providing a variational upper bound for the ground-state energy E:

$$E = min_{\vec{\theta}} \langle \Psi_0 | U_{\vec{\theta}}^\dagger H U_{\vec{\theta}} | \Psi_0 \rangle \qquad (7)$$

Classical optimizers are essential in hybrid algorithms like VQE, where they iteratively adjust circuit parameters to minimize energy despite hardware noise and limited coherence. Their performance directly affects convergence speed, stability, and accuracy, making careful selection critical for efficient and reliable quantum simulations.

The systematic approach described above combines quantum and classical computation, enabling us to explore the electronic structure of molecular systems even with the limitations of current quantum hardware. Our approach builds on recent efforts to study complex surface and defect-related phenomena using advanced electronic structure techniques. For instance, Gujarati et al.[28] demonstrated quantum computation of surface reactions using classical local embedding techniques combined with quantum solvers to manage computational costs. Similarly, Nieman et al.[41] used high-level multireference methods to investigate the electronic structure of single-vacancy defects in graphene analogues, though entirely on classical hardware. Both studies underscore the importance of multireference treatment and local embedding in modeling realistic systems and this work extends those ideas by implementing a hybrid quantum-classical framework that integrates MCSCF with quantum variational solvers (via VQE), exemplifying the type of approach envisioned in recent reviews[42] to address complex problems using currently available quantum resources.

**Potential Energy Surface (PES) of water dissociation using VQE**

To ensure our framework is robust and effective for studying materials and molecules with strong correlation effects, we first validated it using a simpler test case: the dissociation of a water molecule in a vacuum. This preliminary study served as a foundation for systematically evaluating and optimizing key control parameters, such as the choice of ansatz, the size of the active space, classical optimization methods, and qubit mapping strategies. These optimizations were critical in refining our approach to more complex systems exhibiting strong electronic correlations.

The PES for water dissociation is calculated by determining the ground-state energies of the H–O–H system as the O–H bond length is varied (inset of Fig. 2(a)). These calculations are performed using the hybrid quantum-classical framework MCSCF-VQE described earlier. Figure 2(a) illustrates the PES results obtained using different ansatz types: PUCCD[43], SUCCD[43], and UCCSD[44]. Please refer to the methods section for additional information on what these ansatze are. The size of the active space chosen when varying the ansatz is 6 electrons and 4 spatial orbitals (6e4o) to balance the computational cost and speed of the simulations. We observe that PUCCD produces the most accurate PES but is only valid for systems with no unpaired electrons. In addition, the computation times for each ansatz, shown in Fig. 2(b), are comparable across all three configurations, presumably because circuit depths didn't differ drastically for the small system.

Earlier, we mentioned that classical optimizers facilitate the refinement of ansatz parameters, ensuring an optimal representation of strongly correlated electronic states in quantum simulations. Figure 2(c) explores the impact on the PES calculations due to varying the classical optimizer—COBYLA[45], SLSQP[46], SPSA[47] and L-BFGS-B[48]. Details about the characteristics of each of these optimizers are mentioned in the Methods section. We observe that the COBYLA optimizer not only predicts the correct PES but also achieves the shortest execution time, as depicted in Fig. 2(d). This makes COBYLA the preferred optimizer for subsequent simulations. One caveat to mention here is that SPSA is more robust towards noise, hence more applicable for current era quantum hardware, but also takes longer to execute.

Figures 2(e) and 2(f) examine the effect of varying the number of electrons in the active space, which defines the second-quantized Hamiltonian. Adding more electrons does not significantly affect the PES but increases computation time, due to the additional qubits required for a larger active space. These findings emphasize the need to carefully balance active space size and computational resources to optimize simulation accuracy and efficiency.

**Incorporation of error mitigation to improve noise-augmented quantum computations**

To improve quantum simulation accuracy, we applied matrix-free measurement mitigation (M3)[49], a scalable technique that reduces readout errors. Figure 3(a) compares the PES of water dissociation, calculated via MCSCF-VQE simulations, to the ideal PES obtained through exact diagonalization of the Hamiltonian simulated on a noiseless backend. M3 slightly improved energy predictions for stretched water geometries in our MCSCF-VQE simulations. The constraints of existing quantum hardware introduce significant susceptibility to environmental fluctuations, leading to hardware-induced errors that impact computational accuracy. To model these realistic conditions, we introduced noise via an empirical model. As shown in Fig. 3b, unmitigated simulations deviated significantly from exact results, while M3 effectively corrected these errors—highlighting the importance of error mitigation for reliable quantum chemistry in the NISQ era. Fig. 3c and Fig. 3d plot the absolute energy errors relative to the ideal curves for the noiseless and noise-augmented backend simulators respectively.

**Benchmarking on benzene: CCSD with MCSCF-VQE**

Graphene–metal interactions exhibit strong correlation effects, particularly when significant charge transfer occurs. DFT studies show that GGA[52] performs reliably for metal–molecule systems only when the difference between work function (W) and electron affinity ($E_a$) exceeds 7 eV. For pristine and single-vacancy graphene (W ≈ 4.3–4.49 eV) interacting with Fe, Co, and Ni ($E_a$ < 1.2 eV), this criterion is not met, indicating that GGA is insufficient. These systems also display multireference character due to near-degenerate states and bond rearrangements, making them ideal testbeds for quantum algorithms targeting strongly correlated regimes, with relevance to catalysis and quantum sensing.

Due to the large number of electrons in metal-coronene (39 atoms), post-Hartree-Fock methods are computationally prohibitive on classical computers. Instead, we first employ the MCSCF-VQE method to validate water and metal adsorption on benzene—an aromatic prototype for graphene where dispersion and long-range correlation dominate.

Given the hardware constraints of NISQ devices, careful selection of a compact yet representative active space is essential. We employed density difference and natural orbital analyses—commonly used in surface reactions[28]—to guide this selection. For efficiency, the active space was limited to 6 spatial orbitals. Initial DFT calculations using the PBE functional informed the electronic configuration, which was further refined via local density of states to identify key orbitals. For water adsorption on benzene and coronene, the active space was defined as CAS(8e,6o), comprising the oxygen lone pair orbitals (sp³ hybridized), hydrogen 1s orbitals, and the $2p_z$ orbitals of the nearest carbon atoms.

Binding energy of metal M (or water) is calculated from the following expression:

$$E_B^M = E_{M:gr} - E_M - E_{gr} \qquad (3)$$

where, $E_{M:gr}, E_M, E_{gr}$ denote the electronic energies of metal-coronene complex, metal, and coronene, respectively.

For benzene–H$_2$O (Table 1), the MCSCF-VQE binding energy (−0.163±0.0384 eV) is within 0.015 eV of the CCSD benchmark (−0.149 eV), well inside chemical accuracy and the VQE run-to-run variance, while HF underbinds (−0.088 eV) and GGA-DFT overbinds (−0.178 eV). This strong agreement highlights the accuracy and reliability of our hybrid method in capturing weak, correlation-driven interactions in extended pi-conjugated systems—demonstrating near-CCSD performance for systems that are otherwise intractable with conventional post-HF methods. Our results align well with prior CC benchmarks, validating the framework's ability to model weak adsorption in 2D carbon-based materials[50, 51].

We then stress-tested the same framework on open-shell Fe/Co/Ni adsorption on benzene, retaining UCCSD to accommodate spin flexibility (Fig. 4b). Relative to CCSD (Fe −1.104 eV, Co −1.006 eV, Ni −0.661 eV), MCSCF-VQE gives −0.998, −0.891, and −0.855 eV, i.e., absolute errors of 0.106, 0.115, and 0.194 eV respectively. By contrast, GGA-DFT over-binds by about 1 eV on average (−2.255, −1.817, −2.458 eV), while HF spuriously yields repulsive binding energies (+1.412, +2.006, +1.477 eV). The ordering and magnitudes from MCSCF-VQE also track CCSD (|Fe| > |Co| > |Ni|), with the largest deviation on Ni where MCSCF-VQE predicts a stronger interaction between benzene and metal. Taken together with the water case, these results show that a small, physically motivated active space plus a spin-adaptive ansatz can reproduce near-CCSD energetics for both dispersion- and multireference-like adsorption on benzene.

| Binding Energies (eV) | Benzene | | | |
|---|---|---|---|---|
| | Water | Fe | Co | Ni |
| HF | -0.088 | 1.412 | 2.006 | 1.477 |
| DFT | -0.178 | -2.255 | -1.817 | -2.458 |
| CCSD | -0.149 | -1.104 | -1.006 | -0.661 |
| MCSCF-VQE | -0.163 | -0.998 | -0.891 | -0.855 |

Table 1: Comparison of binding energies (eV) between benzene and transition metals/water obtained from MCSCF-VQE, CCSD, DFT and HF. The standard deviation for binding energies for water is 0.0384 eV.

**Predicting binding energies on pristine and single vacancy coronene**

Scaling to larger graphene analogues makes conventional post-HF references impractical, so we retain the MCSCF-VQE strategy on coronene (C$_{24}$H$_{12}$) and single-vacancy coronene (V$_C$). The active space consists of 6 spatial orbitals and was selected based on orbitals contributing significantly to metal–coronene bonding (metal 3d/4s and proximal carbon 2p orbitals), as identified through electronic structure analysis in Fig. 5a. Two clear patterns emerge from the data, as observed from Table 2 and Fig. 5b.

First, water remains a weak physisorbate: the MCSCF-VQE binding changes only slightly between coronene and V$_C$ (−0.121 → −0.108 eV; Δ ≈ +0.013 eV), mirroring its dispersion-dominated character and the limited scope for charge transfer at either site. Second, transition-metal binding is highly site-sensitive. On pristine coronene, MCSCF-VQE finds moderate chemisorption (Fe −1.850 eV, Co −1.832 eV, Ni −1.765 eV), while DFT even misclassifies Fe and Co as unbound (+0.547 and +0.907 eV) and substantially underestimates Ni's binding (−0.547 eV versus −1.765 eV). Introducing a single carbon vacancy drives a dramatic increase in the MCSCF-VQE binding energies—Fe strengthens to −8.609 eV, Co to −7.507 eV, and Ni to −6.157 eV—consistent with the vacancy's electron-deficient, highly reactive character. In this strongly bound regime, DFT again tends to overbind relative to MCSCF-VQE (e.g., Fe −8.852 eV, Co −8.656 eV, Ni −7.878 eV), echoing known limitations of semi-local functionals for metal–defect interactions in $sp^2$ carbon.

These coronene trends are chemically intuitive: the vacancy perturbs the pi-conjugated framework by depleting electron density and generating under-coordinated carbon atoms. This induces localized high-energy states near the Fermi level and effectively lowers the local electron affinity threshold, thereby increasing the likelihood of charge transfer from adsorbates. Further, the trend in metal adsorption energies on pristine and defective coronene (Fig. 5b) reflects the differing electron affinities ($E_a$) of the transition metals and their ability to engage in charge transfer with the electron-deficient carbon vacancy. Fe, with the lowest $E_a$ among the three metals, donates electrons more readily to the vacancy site, resulting in the strongest binding (-8.609 eV). In contrast, Ni, which has the highest $E_a$, exhibits weaker interaction (-6.157 eV), consistent with its reduced electron-donating character. These results highlight the framework's ability to resolve subtle correlation-driven interactions that are often critical in catalytic and sensing applications, particularly where defect engineering modulates reactivity.

| Binding Energies (eV) | Coronene | | $V_C$ - coronene | |
|---|---|---|---|---|
| | DFT | MCSCF-VQE | DFT | MCSCF-VQE |
| $H_2O$ | -0.175 | -0.121 | -0.167 | -0.108 |
| Fe | 0.547 | -1.850 | -8.852 | -8.609 |
| Co | 0.907 | -1.832 | -8.656 | -7.507 |
| Ni | -0.547 | -1.765 | -7.878 | -6.157 |

Table 2: Comparison of binding energies (eV) for pristine and single-vacancy coronene with metals/water obtained from MCSCF-VQE and DFT. DFT consistently over predicts the binding energies and spuriously predicts positive binding energies for Fe and Co on pristine coronene.

**Estimate of computational resources required**

In terms of computational effort, moving from the water benchmark (4–6 spatial orbitals) to the coronene-metal systems (6 spatial orbitals with 8 electrons) resulted in a noticeable increase in simulation complexity. The number of Pauli terms in the qubit Hamiltonian rose from ~200 to over 600, and the UCCSD circuits required approximately 300–400 two-qubit gates for the larger cases. Despite this, circuit depths and shot counts remained within the capabilities of current high-performance quantum simulators, confirming that these strongly correlated systems are tractable on NISQ-era platforms.

**Conclusion**

In conclusion, we present a hybrid quantum-classical framework that combines MCSCF and VQE to accurately model strongly correlated systems beyond the reach of conventional methods. Initial benchmarking on water dissociation enabled systematic tuning of key parameters—including ansatz selection, active space design, and error mitigation—ensuring robust performance. Applying this approach to water adsorption on graphene analogues yielded binding energies in close agreement with high-level post-Hartree-Fock results, validating the framework's ability to capture correlation-driven interactions. We further extended the method to study transition metal adsorption on pristine and defective graphene analogues, successfully capturing charge transfer and multireference effects often misrepresented by standard DFT. While classical Full CI is tractable for the modest active spaces used here, our primary goal is to establish a scalable methodology for systems where classical CASCI becomes intractable. This study serves as a prototype demonstration, laying essential groundwork for applying quantum solvers to increasingly complex problems as quantum hardware evolves. The framework's accuracy, adaptability, and forward scalability make it a promising tool for quantum simulations in catalysis, sensing, and quantum materials design in the NISQ era and beyond.

**Methods**

**Geometry optimization and analysis of electronic configuration**

DFT was used to optimize the geometries used in binding energy simulations These simulations are performed using the Vienna Ab initio Simulation Package (VASP)[53, 54] with the interactions between nuclei and electrons defined by projector augmented wave (PAW)[55] method. The Perdew–Burke-Ernzerhof (PBE)[52] functional is used for exchange correlation energy under the Generalized Gradient Approximation (GGA)[52]. A 2×2×2 grid has been used in these simulations along with an energy cut-off of 500 eV and the simulation box size used was 28×28×28 Å$^3$. The same parameters are also used to calculate the electronic configurations and local density of states of the systems studied.

**Multiconfigurational self-consistent field (MCSCF) and CCSD in PYSCF**

In this work, we use the hybrid quantum-classical method MCSCF-VQE to perform electronic energy calculations to describe the potential energy surface of water and binding energies of transition metals interacting with graphene-analogues like coronene. The hybrid method consists of fragmenting the system into two parts-- active space and environment. Mean-field calculations for the system are performed at the level of Hartree-Fock or Density-Functional Theory using the PySCF[56, 57] package. The basis set used is STO-6G for these systems. Once the mean-field calculations have been performed, the active space is chosen depending on the type of interaction we are trying to study, and the effective Hamiltonian for the active space is generated using the CASCI function in PYSCF. These mean field calculations also serve as initial values for the VQE calculations performed for the active space at the level of CCSD. The MCSCF-VQE framework is compared with the PYSCF driver inbuilt to Qiskit to check its validity as shown in Fig. SI2. This underscores the importance of validating our framework against reference implementations for small, tractable systems. For a minimal test case, our VQE approach successfully reproduced the exact CASCI energy within chemical accuracy, confirming the correctness of the quantum-classical workflow and its suitability for scaling to more complex systems.

**VQE: UCCSD Ansatz and classical optimizers**

VQE is run using the Qiskit SDK[61] using its high-performance quantum computing simulator with realistic noise models called Aer[61]. Using this simulator also enables parallelization of the code using CPUs/GPUs to estimate energies and other electronic properties. As mentioned in the earlier sections, variations of the UCCSD ansatz are used to generate the parametrized wavefunction, which are obtained by applying the exponential of the anti-Hermitian operator $T - T^\dagger$ to the Hartree-Fock state. $T$ is a linear combination of single and double excitations from occupied (indexed as p, q) to virtual spin-orbitals (indexed as a,b). More specifically,

$$|\psi_{\text{UCCSD}}(\theta)\rangle = e^{T-T^\dagger}|\psi_{\text{HF}}\rangle \quad (4)$$
$$T = T_1 + T_2$$
$$T_1 = \sum_{aj}(\theta^R_{ap} + i\theta^I_{ap})\hat{a}^\dagger_a \hat{a}_p$$
$$T_2 = \sum_{apbq}(\theta^R_{apbq} + i\theta^I_{apbq})\hat{a}^\dagger_a \hat{a}^\dagger_b \hat{a}_q \hat{a}_p$$

where $\hat{a}^\dagger_a/\hat{a}_p$ creates/destroys an electron at spin-orbital a/p, and the coefficients $\theta = \{\theta^R_{ap}, \theta^I_{ap}, \theta^R_{apbq} \text{ and } \theta^I_{apbq}\}$ are variational parameters

The ansatze used are subclasses of the Unitary Coupled Cluster (UCC) framework. UCCSD includes both single and double excitations, while SUCCD and PUCCD include only doubles. SUCCD enforces spin symmetry under particle exchange, whereas PUCCD restricts excitations to occur in parallel across α and β spin orbitals.

Classical optimizers fall into gradient-based (L-BFGS-B[48], SLSQP[46]) and gradient-free (COBYLA[45], SPSA[47]) categories. Gradient-based methods perform well under low-noise conditions but are sensitive to complex energy landscapes. In contrast, COBYLA and SPSA are more robust in noisy environments, requiring fewer quantum evaluations and offering better performance on NISQ hardware. The optimal optimizer choice depends on problem complexity and device noise. Convergence characteristics are detailed in SI1 in the supplementary material.

Given that optimizer convergence is often a critical challenge in VQE, we evaluated the performance of our selected optimizers for the largest active space considered (8e/6o). For these cases, the COBYLA optimizer consistently converged within ~50–70 iterations, depending on the system, without encountering significant issues such as local minima or oscillatory behavior. To further ensure robustness, we performed multiple VQE runs with randomized initial parameter guesses; all runs converged to near-identical energies within the expected error margins, suggesting successful convergence to the global minimum. These observations reinforce the reliability of the optimization strategy used in our simulations.

## Acknowledgements

This research was supported by the National Science Foundation (NSF) under Grant # CMMI-2527151. We also acknowledge the Texas Advanced Computing Center (TACC) at The University of Texas at Austin for providing the computing resources on Lonestar6 a under the allocation No. DMR22008.

## Data Availability

The data that supports the findings of this study are available within the article and its supplementary material.

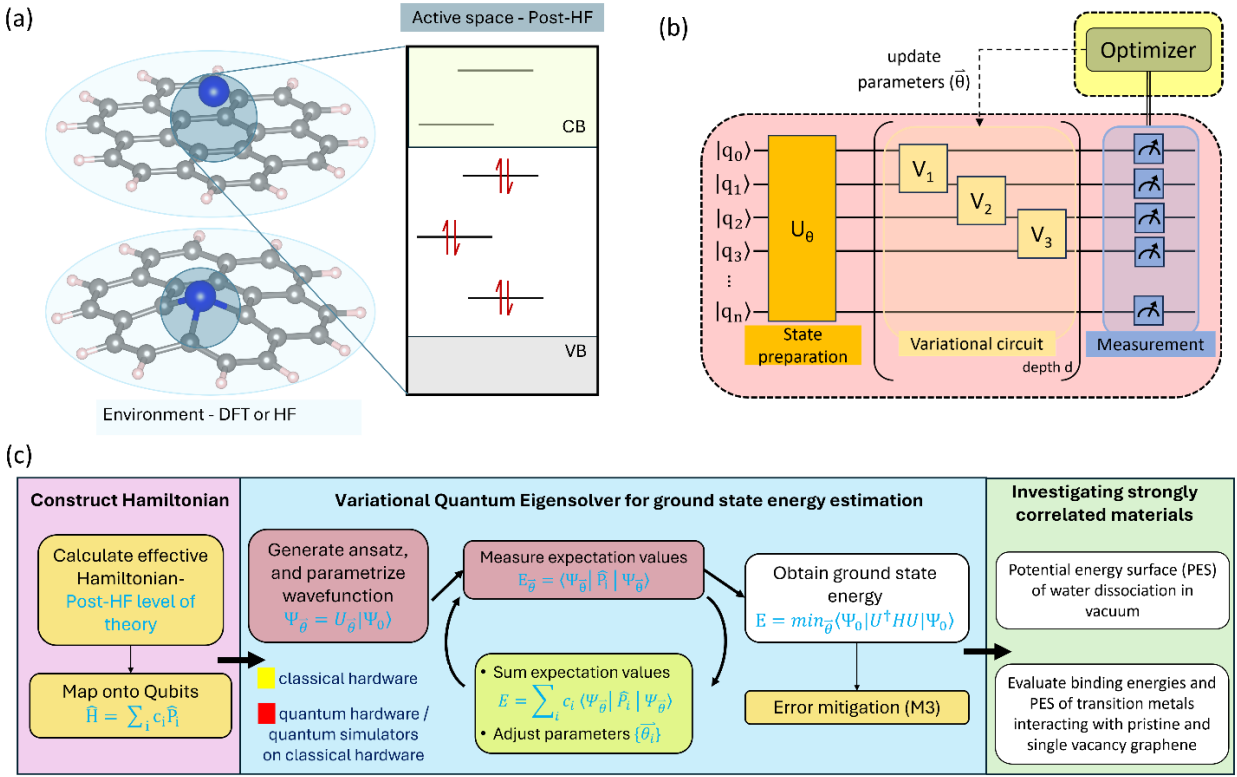

**Figure 1. Framework of evaluating electronic properties of a strongly correlated material and adsorbent system where electron correlations can be significant**. (a). An example of such a system is shown consisting of a transition metal adsorbed on pristine and defective coronene. The environment is evaluated at a mean-field level of theory like HF or DFT. The effective Hamiltonian of the active space is evaluated using a chosen post-HF method (b). Representative quantum circuit for execution of VQE. (c). Workflow of the VQE algorithm for estimating mechanical and electronic properties of strongly correlated materials. The yellow and red shaded modules in (b), (c), indicate computations done on classical hardware and quantum hardware (or quantum simulators performed using classical hardware) respectively.

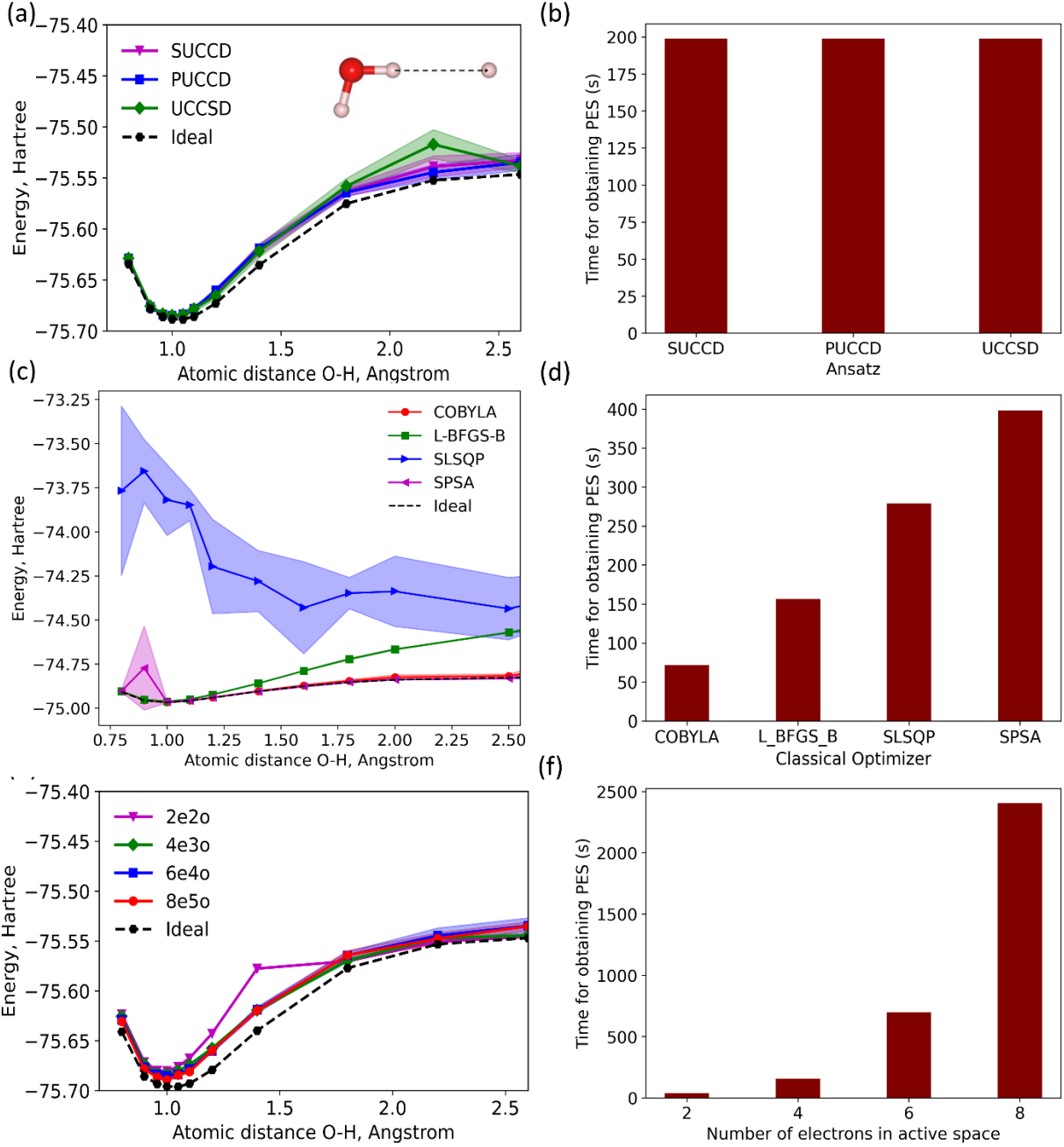

**Figure 2: Potential Energy Surface (PES) of water dissociation using VQE.** (a), (c), (e) Comparison of potential energy surfaces of water varying ansatz (6e4o), classical optimizer (6e4o), and number of electrons in the active space respectively. The legend indicates the number of electrons 'e' and number of spatial orbitals 'o' used in the simulation. (b), (d), (f) Times required to obtain the PES of water dissociation by varying the ansatz, classical optimizer and number of electrons in the active space.

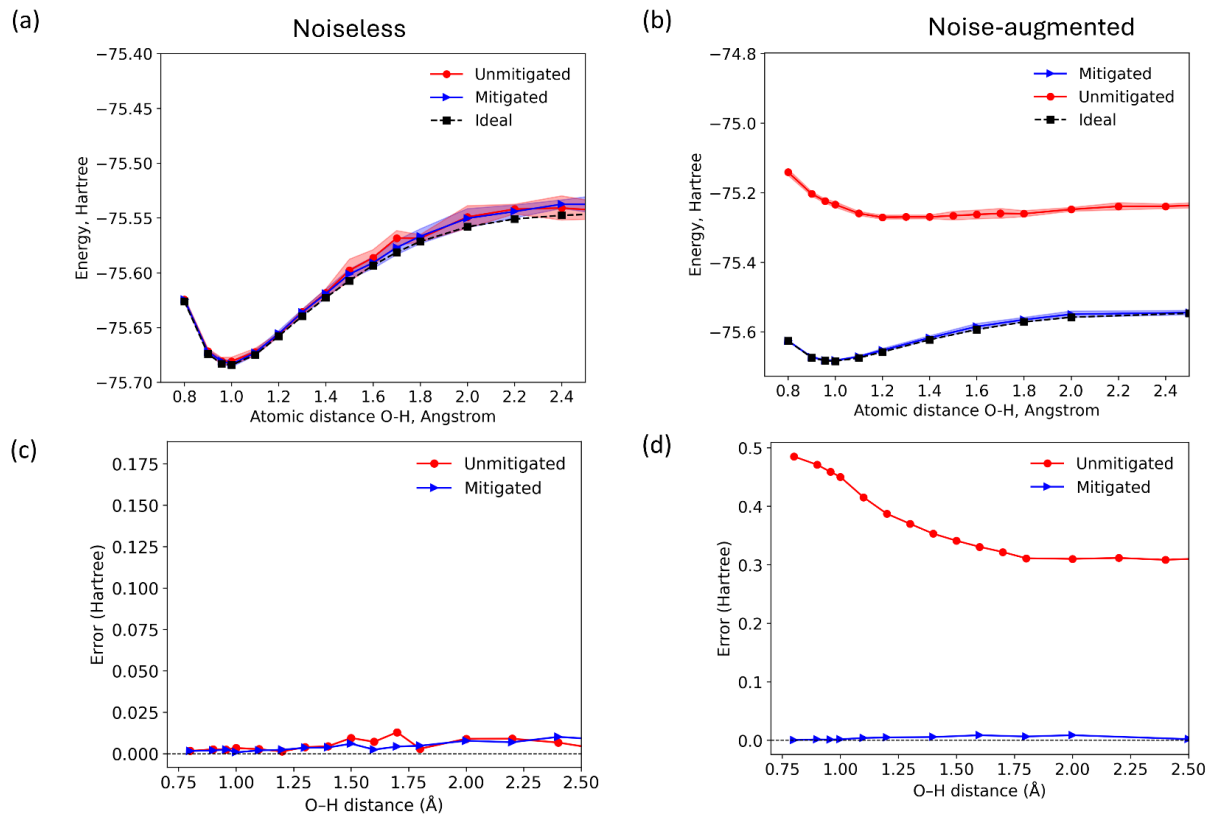

**Figure 3. Error mitigation of water dissociation using VQE and MCSCF-VQE.** (a) Comparison of the mitigated, unmitigated and ideal PES curves of water dissociation with the Noiseless backend. (b) Comparison of the mitigated, unmitigated and ideal PES curves of water dissociation with the Noise-augmented backend. Incorporation of noise causes large discrepancies in the predicted PES (unmitigated; green circles), which is improved by mitigating error using M3 (mitigated; blue triangles) (c) Absolute energy error relative to the ideal curve for the Noiseless backend. Both unmitigated and M3-mitigated results track the ideal within a few millihartree across the O–H scan, consistent with panel (a). (d) Absolute energy error relative to the ideal curve for the Noise-augmented backend. Hardware-like noise induces large systematic deviations (unmitigated; red circles, up to ~0.5 Ha) that are suppressed by M3 (mitigated; blue triangles) to near-zero across all O–H separations.

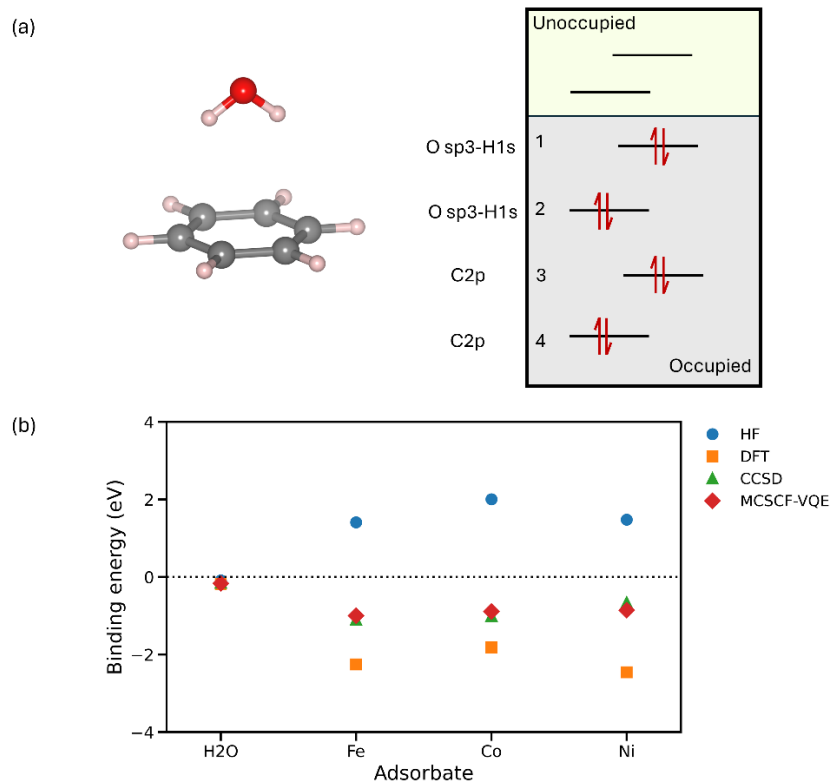

**Figure 4: Water (2-leg orientation) interacting with benzene. (a)** The electronic configuration of the active space is shown, along with the orbitals which contribute most significantly to the electron density. Oxygen is shown in red, carbon in teal and hydrogen in white. **(b)** Binding energies (eV) for $H_2O$ and 3d transition metals on benzene computed with HF (circles), DFT (squares), CCSD (triangles), and MCSCF-VQE (diamonds). The dotted line marks 0 eV; negative values indicate exothermic adsorption. For water, all methods give weak physisorption near the CCSD value (CCSD −0.149 eV; MCSCF-VQE −0.163 eV). For the metals, HF predicts spurious repulsion, while DFT overbinds relative to CCSD. MCSCF-VQE closely tracks CCSD with absolute errors of ≈0.11 eV (Fe), 0.12 eV (Co), and 0.19 eV (Ni), recovering the trend |Fe| > |Co| > |Ni|.

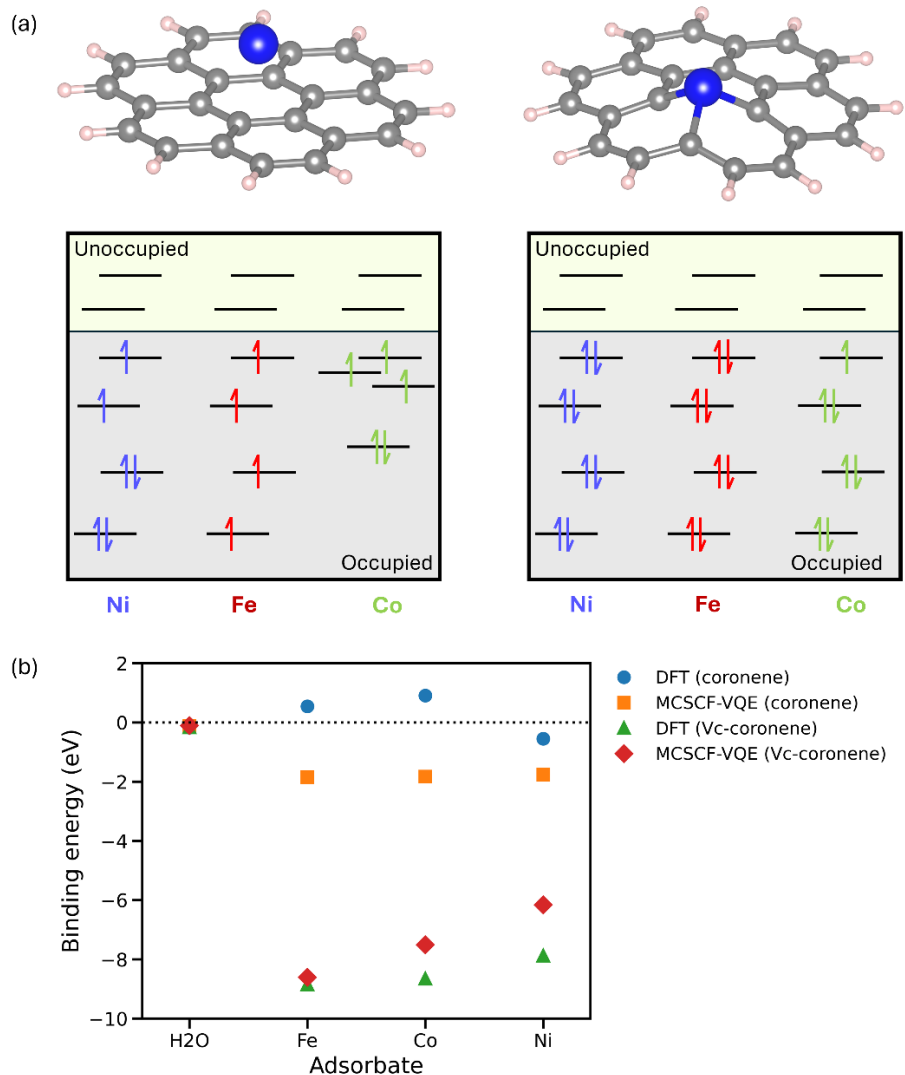

**Figure 5: Fe, Co, Ni interacting with pristine and single vacancy coronene. (a)** For each of these systems, the electronic configuration of the active space is shown, along with the orbitals which contribute most significantly to the electron density. Metal is shown in blue, carbon in teal and hydrogen in white. **(b)** Binding energies (eV) for $H_2O$ and 3d transition metals on pristine coronene (DFT—circles; MCSCF-VQE—squares) and on single-vacancy coronene ($V_C$) (DFT—triangles; MCSCF-VQE—diamonds). The dotted line marks 0 eV; negative values denote exothermic adsorption. Water remains weakly physisorbed on both substrates, whereas Fe/Co/Ni bind moderately on pristine coronene; note that DFT incorrectly predicts Fe and Co to be unbound on pristine coronene, while MCSCF-VQE yields ∼−1.8 eV. Introduction of a single carbon vacancy dramatically strengthens metal binding (MCSCF-VQE: Fe = −8.609 eV, Co = −7.507 eV, Ni = −6.157 eV) with little effect on water, highlighting the strong defect-induced charge-transfer/covalency in the metal–$V_C$ system.

# Supplementary material

**Hybrid Quantum-Classical Simulations of Graphene Analogues: Adsorption Energetics Beyond DFT**

Archith Rayabharam[1], N. R. Aluru[1,*]

[1] Walker Department of Mechanical Engineering, Oden Institute for Computational Engineering and Sciences, The University of Texas at Austin, Austin, Texas, USA

*Corresponding author, email: aluru@utexas.edu

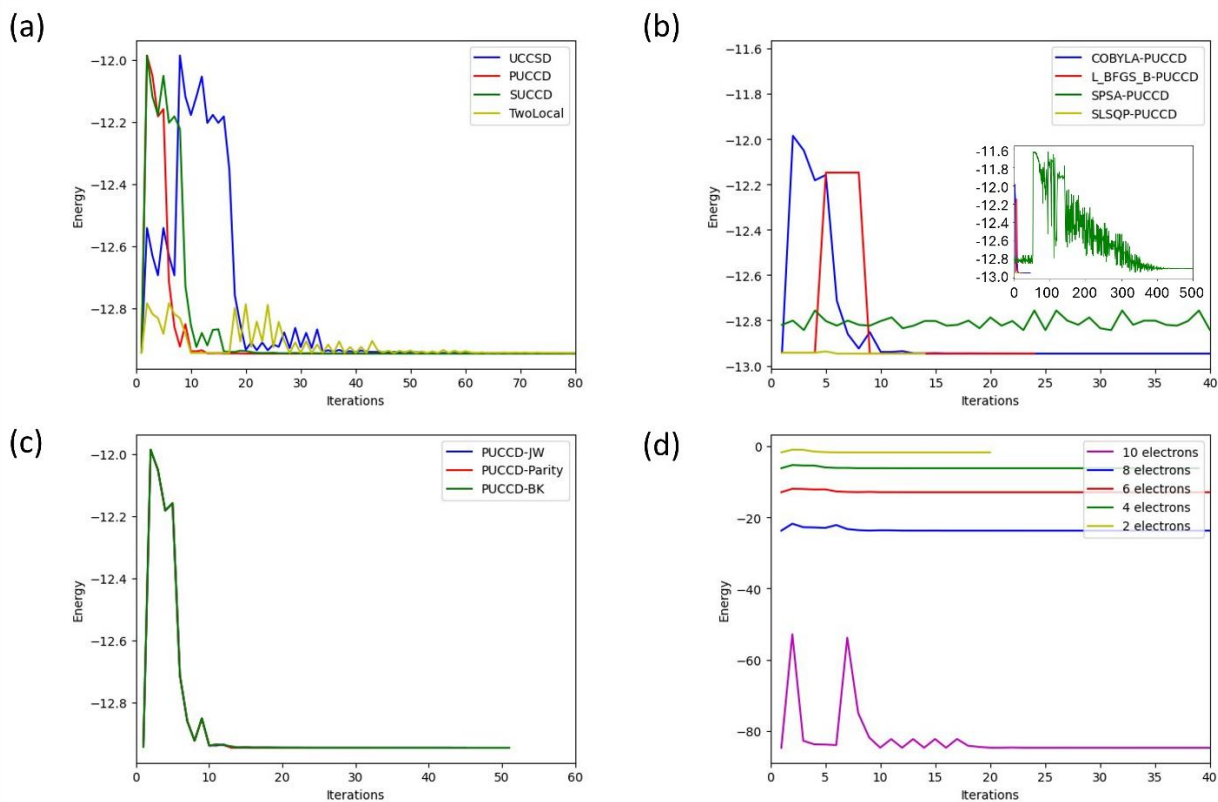

**Figure SI1: Convergence characteristics of water dissociation using VQE.** (a) For ansatz, PUCCD works best and converges within 20 iterations. Further, the TwoLocal ansatz starts off with a better estimate of energy but converges the slowest. (b) For classical optimizers, SLSQP (least squares) works best, and converges within 10 iterations. SPSA (Simultaneous Perturbation Stochastic Algorithm) takes around 400 iterations to converge, making it the worse by far for noiseless simulators. (c) Type of mapping does not affect the convergence characteristics. (d) Having more electrons in the active space increases the number of iterations required to converge.

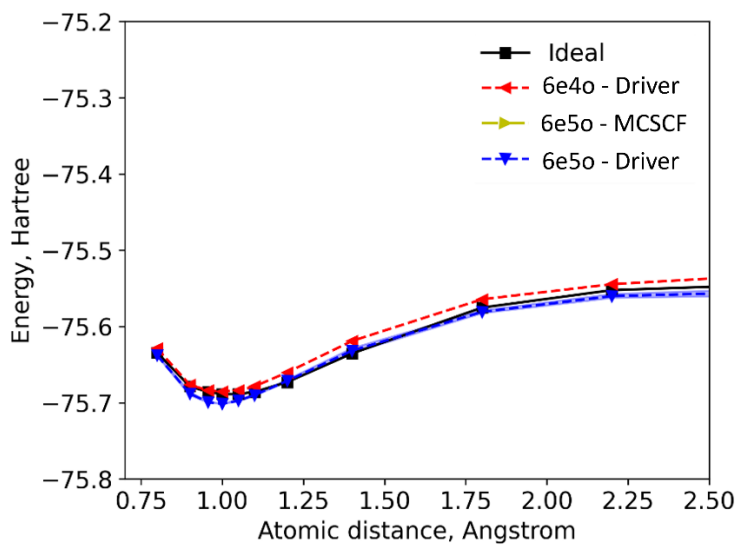

**Figure SI2: Validation of the MCSCF-VQE framework with the PYSCF driver:** PES of water dissociation with no noise. The driver allows for quick analysis for electronic systems using quantum computers and is built into Qiskit. Although, it does not offer much flexibility in investigation of customized systems, it is used to validate the MCSCF-VQE framework used in the study. The framework is specifically validated for water before being extended to interactions between graphene-analogues and water, transition metals.

## Optimization using CASSCF, and ADAPT-VQE

In this work, orbitals for the active space are taken from single-determinant DFT or HF and are not fully relaxed, as in true CASSCF. However, such unoptimized orbitals may be suboptimal for multireference systems. Recent orbital-optimized VQE approaches emulate self-consistent CASSCF-like relaxation on quantum hardware, improving accuracy and noise resilience[1-3]. Incorporating these methods is a promising direction for future improvements.

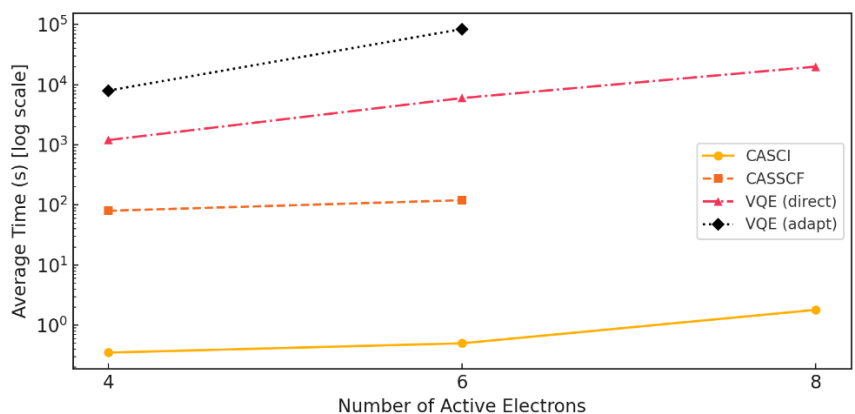

**Figure SI3: Average wall-clock time (seconds, log scale) versus number of active electrons for four solvers**: CASCI (orange, solid), CASSCF (orange, dashed), VQE (direct UCCSD) (magenta, dash-dot), and ADAPT-VQE (black, dotted). Runtime grows with active-space size; ADAPT-VQE is the most expensive due to iterative operator-pool growth and repeated measurements, VQE (direct) is intermediate, and classical CASCI/CASSCF are fastest at these sizes. All timings were obtained on the same hardware and active-orbital space.